\documentstyle[epsf]{article}
\begin{document}
\title{\textbf{Comparison of $\Lambda$ and $\Sigma^0$ threshold production 
       in proton-proton collisions}}
\author{}
\maketitle
\begin{center}S.~Sewerin$^1$,
 G.~Schepers$^{1, 2}$,
 J.T.~Balewski$^{1, 3, ^*}$,
 A.~Budzanowski$^3$,
 W.~Eyrich$^4$,
 M.~Fritsch$^4$,
 C.~Goodman$^5$,
 D.~Grzonka$^1$,
 J.~Haidenbauer$^1$,
 C.~Hanhart$^1$,
 M.~Hofmann$^1$,
 L.~Jarczyk$^6$,
 M.~Jochmann$^7$,
 A.~Khoukaz$^2$,
 K.~Kilian$^1$,
 M.~K\"ohler$^7$,
 T.~Lister$^2$,
 P.~Moskal$^{1, 6}$,
 W.~Oelert$^1$,
 I.~Pellmann$^1$,
 C.~Quentmeier$^2$,
 R.~Santo$^2$,
 U.~Seddik$^8$,
 T.~Sefzick$^1$,
 J.~Smyrski$^6$,
 F.~Stinzing$^4$,
 A.~Strza\l kowski$^6$,
 C.~Wilkin$^9$,
 M.~Wolke$^1$,
 P.~W\"ustner$^7$,
 D.~Wyrwa$^{1, 6}$
\end{center}
\noindent
$^1$ IKP,  Forschungszentrum J\"ulich, D-52425 J\"ulich, Germany \\
$^2$ IKP, Westf\"alische Wilhelms--Universit\"at, D-48149 M\"unster, Germany \\
$^3$ Institute of Nuclear Physics, PL-31-342 Cracow, Poland \\
$^4$ IP, Universit\"at Erlangen/N\"urnberg, D-91058 Erlangen, Germany \\
$^5$ IUCF, Bloomington, Indiana, IN 47405, USA \\
$^6$ Institute of Physics, Jagellonian University, PL-3059 Cracow, Poland \\
$^7$ ZEL,  Forschungszentrum J\"ulich, D-52425 J\"ulich,  Germany \\
$^8$ NRC, Atomic Energy Authority, 13759 Cairo, Egypt \\
$^9$ University College London, London WC1E 6BT, United Kingdom \\
\newpage
\begin{flushleft}
Threshold measurements of the associated strangeness production reactions
$pp \to p K^+ \Lambda $ and $pp \to p K^+ \Sigma^0 $ are presented. 
Although slight differences in the shapes of the excitation functions are 
observed, the most remarkable feature of the data is that at the same excess 
energy the total cross section for the $\Sigma^0 $ production appears to be 
{\bf{about}} a factor of $28$ smaller than the one for the $\Lambda$ particle.
It is concluded that strong $\Sigma^0p$ final state interactions, and in 
particular the $\Sigma N \to \Lambda p $ conversion reaction, are the likely 
cause of the depletion for the yield in the $\Sigma$ signal. This hypothesis 
is in line with other experimental evidence in the literature.  
\end{flushleft}
$^*$ present address IUCF, Bloomington, Indiana, USA \\
\\
\\
\newpage
At COSY - J\"ulich the ``COSY-11'' collaboration  studies
the production of strangeness in proton-proton scattering at threshold,
using an internal cluster target facility~\cite{Dom}.
Here we report on experimental data for the reactions:
$pp \rightarrow pK^+\Lambda$ and $pp \rightarrow pK^+\Sigma^0$ at excess
energies \mbox{$Q = \sqrt s - m_p -m_{K^+} - m_{\Lambda(\Sigma)} \le 12.9$~MeV}.
In this region low partial waves are expected to dominate.

In the non-perturbative domain of COSY energies, and especially in the
threshold region, the physics of strangeness production is most appropriately
described in terms of meson exchange. In such models both strange and
non-strange exchanges with or without intermediate resonance excitation
can occur. In addition to the commonly considered $\pi$ and $K$ exchange
contributions, see e.g. Refs~\cite{Colin96}~-~\cite{Sibi98}, 
the exchange of heavier non-strange and strange 
mesons~\cite{Tsushuma98} and their interference effects might have an 
influence on the strangeness production process.  
Effects of coupling constants, resonances and
final state interaction (FSI) might become visible when comparing
different final states like the $\Lambda p K^+$ and the  $\Sigma^0 p K^+$
channels. For instance, the ratio of the coupling constants
$g^{2}_{\Lambda N K}/g^{2}_{\Sigma N K}$, as extracted from different
reactions involving hyperons, varies between 0.08 and 27
\cite{Li98} - \cite{Sibi98}, \cite{Deloff} - \cite{Holzenkamp}.
As a consequence, model predictions of $\Lambda$ and $\Sigma^0$ production 
cross sections differ significantly.

The COSY-11 facility, described in detail in Ref.~\cite{Brau96}, allows one 
to measure the four-momenta of the proton and $K^+$ directly, leaving 
the uncharged hyperon to be identified using the missing mass method. 
A regular COSY-dipole magnet, placed downstream of the target, separates the
reaction products with momenta different from that of the circulating beam. 
Two drift chambers detect the directions of the particles and enable a 
momentum determination by ray tracing back through the known magnetic field 
into the target. Particle identification is then performed by measuring the 
time of flight between start and stop scintillators.

The main goal of the present investigation was to measure the $\Sigma^0 $
production and to compare it with $\Lambda$ production near threshold. 
Here we report on the measurement of the total cross section for
$pp\to pK^+\Sigma^0$ at seven energies in the range $3.0 < Q < 12.9$~MeV. 
In the $\Lambda$ case, we have already presented seven data points  from 
$Q=0.68$ to 6.68~MeV~\cite{Bal98}, and we here extend this range also to 
$12.9$~MeV through the addition of three extra points. 
COSY was used  in the ``supercycle'' mode, which allows a repetition 
of a sequence of spills with different parameters from spill 
to spill. In view of the large cross section difference between the production 
of the two hyperons, 10 or 20 spills  at beam momenta equivalent to excess 
energies $Q = 7.7$~MeV, $10.5$~MeV,~and~$12.9$~MeV  above the  $\Sigma^0$
threshold were followed by one spill at the equivalent Q-values above the 
$\Lambda$  threshold. The  spill length was five minutes and the sequence was 
repeated for a total running time of two to three days for each beam momentum. 
The supercycle  
mode compares two similar processes under similar conditions, thus reducing  
possible errors due to shifts in accelerator and/or detector components.

\begin{figure}[ht]
\centerline{
\leavevmode
  \epsfxsize=9.0cm
  \epsfysize=8.5cm
   \hspace{0.0cm}
\epsffile{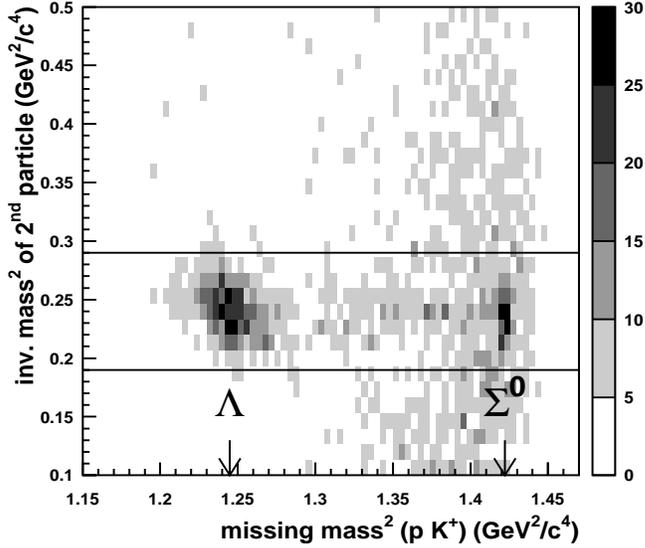}}
   \vspace{0.0cm}
\caption{Square of the mass of the second particle {\it versus} square of the
missing mass, after identification of the first particle as being a proton. 
Data shown are at an excitation energy of $Q_{\Sigma^0} = 12.9$~MeV. The horizontal solid 
lines indicate the limits of the $K^+$-band assumed. The much more populated 
$\pi^+$ and proton bands are off the figure.}
 \label{2dim}
\end{figure}

To isolate hyperon production in the presence of a large background,
all the two-track events, which are candidates for $K^+$ and $p$, were 
selected from the raw data. After the determination of the four-momentum vector 
($E,~\overrightarrow{p}$) of both particles, the missing mass was calculated.
In Fig.~\ref{2dim}, the square of the mass  ($m_{inv}$) of the lighter particle
is plotted {\it versus} the square of the missing mass. A clear $K^+$-band is 
apparent, with enhancements at the positions of the $\Lambda$ and the
$\Sigma^0$ - masses.

\begin{figure}[t]
\centerline{
\leavevmode
  \epsfxsize=8.6cm
  \epsfysize=8.6cm
  \hspace{1.0cm}
\epsffile{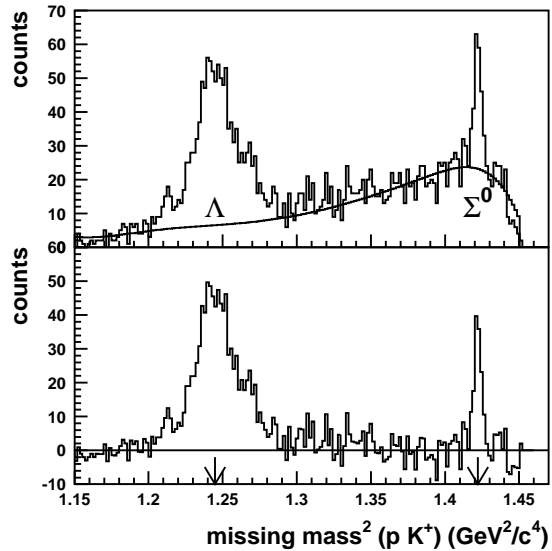}}
  \vspace{0.0cm}
\caption{
Top: Spectrum of missing mass squared for the reaction $pp \rightarrow pK^+ X$
at a beam momentum equivalent to $Q = 12.9$~MeV with respect to the $\Sigma^0$ 
production threshold. The solid line indicates the smoothed background 
distribution obtained by projecting bands above and below the $K^+$-band of 
Fig.~\ref{2dim}.\protect \\
Bottom: Spectrum after background subtraction. 
Equivalent spectra for the measurements close to the $\Lambda$ threshold can be
seen in Ref.~\protect\cite{Bal98}.}
\label{mm2}
\end{figure}

The projection onto the missing mass axis of the band between the
two indicated lines is shown in the upper part of Fig.~\ref{mm2}.
The summed projection of adjacent bands in Fig.~\ref{2dim}
($m^2_{inv}$ = 0.1 -- 0.19~GeV$^2$/c$^4$ and
$m^2_{inv}$ = 0.29 -- 0.34~GeV$^2$/c$^4$ )
is considered as representative of the background spectrum which, after
normalising and smoothing to minimize statistical fluctuations, is shown as 
the solid curve in the figure.
Subtracting this background from the missing mass
spectrum results in the lower part of Figure~\ref{mm2}, which shows clear
$\Lambda$ and $\Sigma^0$ peaks. The sharpness of the latter is a kinematic
effect due to the proximity to the threshold and the same effect is seen in the 
corresponding spectra close to the $\Lambda$ threshold~\cite{Bal98}.

\begin{table}[ht]
\center{
\begin{tabular}[t]{|c|c|c|c|c|c|}
\hline
nominal&extracted &$ p p\rightarrow p K^+ \Sigma^0$  &extracted &$p p\rightarrow p K^+ \Lambda$ & \\[-0.3cm]
Excess energy &excess &cross section  &excess &cross section
&$\displaystyle{\frac{\sigma_{T}(pp \to pK^+ \Lambda )}{\sigma_{T}(pp \to pK^+
\Sigma^0 )}}$ \\[-0.3cm]
$Q$ (MeV)  &energy   &(nb)  &energy &(nb) & \\[0.1cm]
\hline
\hline
   $  3.0 $  &$~2.8$   &$ 1.6  \pm 0.5 $ & & &\\
\hline
   $  5.0 $  &$~5.5$ &$ 5.7  \pm 0.8 $ & & &\\
\hline
   $  7.0 $  &$~7.5$  &$ 8.6  \pm 2.1 $ & & &\\
\hline
   $  7.7 $  &$~8.0$ &$ 9.7  \pm 2.0 $ &$~8.6$  &$ 344  \pm 41 $ &$35 \pm 15$\\
\hline
   $  10.0 $  &$11.1$  &$ 17.5 \pm 3.8 $ & & & \\
\hline
   $  10.5 $  &$10.3$  &$ 12.8 \pm 2.4$ &$10.9$   &$ 385  \pm 27 $ &$30 \pm 9$\\
\hline
   $  12.9 $  &$13.0$  &$ 20.1 \pm 3.0$ &$13.2$   &$ 505  \pm 33 $ &$25 \pm 6$\\
\hline
\end{tabular}
\caption[cross_sigma]{\label{cross_sigma_lambda}
Total cross sections for the $pp \rightarrow pK^+\Sigma^0$ and
$pp \rightarrow pK^+\Lambda$ reactions. Uncertainties of the excess energies are
discussed in the text.}}
\end{table}

Extracted cross sections for both reactions under discussion are listed in 
Table \ref{cross_sigma_lambda}. The luminosity was determined by comparing the 
differential counting rates of elastically scattered protons with data 
obtained by the EDDA collaboration~\cite{EDDA}. The acceptance of the COSY-11 
apparatus was calculated using GEANT Monte-Carlo simulations~\cite{GEANT} with 
a three-body phase-space generator. The errors in the cross sections 
are purely statistical, where
the uncertainty in the $\Sigma^0$ yield is mainly due to the background 
contribution. In addition, there is a total systematic uncertainty of 
$\le 22\%~(\Lambda) $ and $\le 32\%~(\Sigma^0)$, made up of luminosity ($\le 13\%$), 
acceptance ($\pm 4\%$), background subtraction ($\le 15\%$ for $\Sigma^0$ 
and $\le 5\%$ for $\Lambda$ production). In view of the sharp energy 
variation of the cross sections, it is important to verify the 
beam momenta derived using the machine parameters. The estimated accuracy of the COSY
momentum of $0.1\%$ would result in an error of the excess energy of $0.87$~MeV
($0.80$~MeV) at the $\Sigma^0 (\Lambda)$-threshold, respectively. The value of the 
missing mass of the hyperons can be determined from the present data with an 
accuracy of the equivalent excess energy to be $0.4$~MeV. By comparing this with 
the latest compilation~\cite{PDG} one can derive the true excess energy 
within $0.4$~MeV. 

The remarkable feature of the measurements is that at the same value of $Q$
the ratio
\begin{equation}
R_{pp}(Q)=
\frac{\sigma_{T}(pp\to pK^+\Lambda)}{\sigma_{T}(pp\to pK^+\Sigma^0)}
\end{equation}
favours strongly $\Lambda$ over $\Sigma$ production, with an average value
of approximately $28$. 
At much higher energies~\cite{Flam79} the available experimental data indicate
this ratio to be about 2.5, suggesting a strong influence of threshold effects
on the relative $\Lambda$ - $\Sigma^0$ production cross section at low excess
energies.

\begin{figure}[ht]
\centerline{
\leavevmode
  \epsfxsize=9.0cm
  \epsfysize=9.0cm
  \hspace{1.0cm}
\epsffile{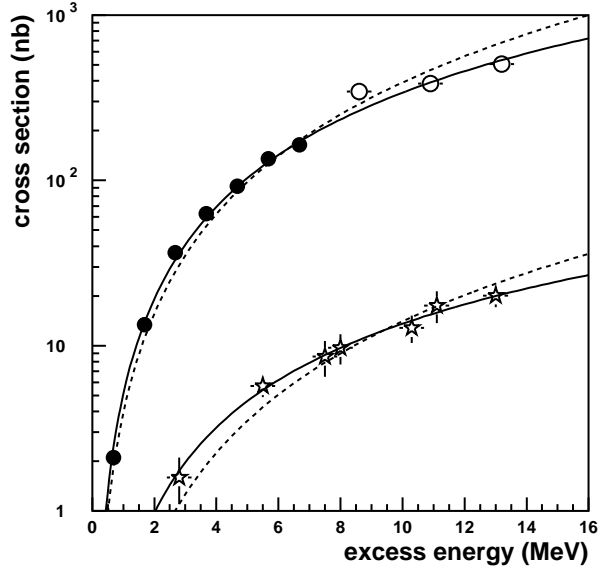}}
  \vspace{0.0cm}
\caption{
 Cross sections for the reactions $pp \rightarrow pK^+\Lambda$ (circles) and
 $pp \rightarrow pK^+\Sigma^0$ (stars). Filled circles represent data published
 in [15].
 Statistical error bars are given or are smaller than the symbol size. The
 horizontal error bars are discussed in the text. The curves represent phase-space fits 
 with proton-hyperon final state interaction (solid curve) and without 
 (dashed line); the latter corresponds to $\epsilon\to\infty$ in eq.~(2).}
\label{cross_section_plus_fit}
\end{figure}

A comparison of the excitation functions for the two reactions in the 
threshold region is shown in Figure~\ref{cross_section_plus_fit}.
If the production is of short range, the energy variation should be determined by
phase-space modified by any final state interaction (FSI). Taking into account
only the dominant hyperon-nucleon FSI, it is expected~\cite{Colin96} that
\begin{equation}
\sigma_T = C\,\frac{Q^2}{\left(1+\sqrt{1+Q/\epsilon}\right)^2}\:,
\end{equation}
where $\epsilon$ represents the energy of a nearby virtual state. A best fit 
to the data is shown in Fig.~\ref{cross_section_plus_fit} and gives:
\begin{eqnarray}
\nonumber
C(\Lambda) = (21.7\pm 1.0)~\mbox{\rm nb/MeV}^2\ \ \ \ \ && \epsilon(\Lambda) =(7.5\pm
1.4)~\mbox{\rm MeV}\:,\\
C(\Sigma^0) = (1.3\pm 0.6)~\mbox{\rm nb/MeV}^2\ \ \ \ \ && \epsilon(\Sigma^0) =(3.1\pm
3.2)~\mbox{\rm MeV}\:.
\end{eqnarray}
This parametrization shows again that the $\Sigma^0$ production yield is much less
than the $\Lambda$ one, with a ratio of the normalisation factors 
being
$C(\Lambda)/C(\Sigma^0) \approx 17$. It is however important to
note that the structure in the $\Sigma^0 p$ FSI is much sharper
than for $\Lambda p$. 

A quantitative explanation of the relatively low $\Sigma^0$ production
cross section observed in $pp$ collisions at threshold has to wait for detailed
theoretical investigations, but already a qualitative discussion can be
presented.

Both $\pi^0$ and $K^+$ exchange diagrams in the $t$-channel can contribute to 
the production of neutral hyperons in $pp$ collisions.
If we consider the one-kaon-exchange contribution alone (and ignore
effects from the hyperon-nucleon final state interaction) then the
$\Lambda / \Sigma^0$ production ratio is essentially given by the
ratio of the coupling constants $g^2_{\Lambda N K}/g^2_{\Sigma NK}$.
There is considerable uncertainty in the values of the
$K^+p\Lambda(\Sigma^0)$ coupling constants
\cite{Li98} - \cite{Sibi98}, \cite{Deloff} - \cite{Holzenkamp}
but, with a suitable choice, it would be possible 
to reproduce the observed large $\Lambda/\Sigma^0$ production ratio in a pure 
kaon-exchange model. It is perhaps fortuitous that the measured ratio happens to 
coincide almost exactly with the value of the SU(6) 
prediction~\cite{deSwart63} for the coupling constant ratio, which is 27/1.  

At the higher beam energy of $2.3$ GeV, corresponding to $Q = 170$~MeV with respect to
the $\Sigma^0$ channel, there has been a detailed inclusive measurement of $K^+$
production in the $pp \to K^+X$ reaction~\cite{Siebert}. 
Significant enhancements are observed at
both the $\Lambda p$ and $\Sigma N$ thresholds with similar magnitudes. Since only
the $K^+$ was detected, there is no way of knowing whether the second rise is due to
true $\Sigma$ production or whether virtually produced $\Sigma$'s are captured on
the nucleon and emerge rather as $\Lambda$'s
through a strong $\Sigma N\to \Lambda p$ final state 
interaction. Such effects are well documented in the literature in, for example,
$K^-$ absorption in deuterium~\cite{Tan1,Tan2,Cline}. Data for fully constrained
$K^-d\to \pi^-\Lambda p$ events with stopping kaons show a steep rise from
threshold with evidence for a strong $\Lambda p$ FSI~\cite{Tan1}. The most
remarkable feature, however, is the  sharp peak at an effective mass of
$m(\Lambda p) = 2129$~MeV/c$^2$,  {\it i.e.}\ at the $\Sigma^0 p$ threshold,
with a FWHM of about 8~MeV/c$^2$.  This is to be associated with the two-step
process $K^-d\to \pi^- (\Sigma N \to \Lambda p)$. Such a very large effect in
deuterium, where the average proton-neutron separation is about 4~fm, 
requires the $\Sigma N $ scattering length to have an 
imaginary part of about 1.4~fm~\cite{Sasha}. This
must lead to much bigger effects in the proton-proton production studied
here, since the large  momentum transfers favour short distances and this in
turn will enhance the $\Sigma^0\to\Lambda$ conversion rate.
Unless the basic physics changes radically between threshold and the energy of the
Saclay measurement, the obvious way to reconcile the two results is to assume that
at COSY-11 many $\Sigma$'s are produced but that most of them are converted 
to $\Lambda$'s in the interaction. In this case one needs a production 
mechanism which offers a larger yield than predicted
by the $K^+$ exchange with SU(6) coupling constants.

In the one-pion-exchange contribution to the threshold 
$pp\to pK^+\Lambda(\Sigma^0)$ amplitudes, the driving terms are proportional to 
the $\pi^0 p\to K^+\Lambda (\Sigma^0)$ amplitudes. There are measurements of 
the $\pi^-p\to K^0\Lambda(\Sigma^0)$~\cite{Jones} and 
$\pi^+p\to K^+\Sigma^+$~\cite{Candlin} cross sections near and somewhat above 
threshold. Assuming isospin $I=\frac{1}{2}$ dominance due to the presence of
the $N^*$(1650), the data of Ref.~\cite{Jones} suggest that near threshold 
\begin{equation}
R_{\pi p}(Q) = \frac{|f(\pi^-p\to K^0\Lambda)|^2}{|f(\pi^-p\to K^0\Sigma^0)|^2}
=\frac{|f(\pi^0p\to K^+\Lambda)|^2}{|f(\pi^0p\to K^+\Sigma^0)|^2}
\approx 0.4\:.
\end{equation}

Neglecting the effects of the hyperon-nucleon final state interactions in a
one-pion-exchange model, one would naively expect
$R_{\pi p}(0) \approx R_{pp}(0)$, which leads to a discrepancy with respect to 
the present data of about two orders of magnitude.  

In reality one expects that both $\pi$ and $K$ exchange
will contribute to the production of hyperons.
In case of $\Lambda$ threshold production the ratio of $K^+$ to $\pi^0$
exchange contributions is roughly given by the ratio of
$|f(K^+p\rightarrow K^+p)|^2$ to $|f(\pi^0 p\rightarrow K^+\Lambda)|^2$. 
Based upon the $K^+p$ ~~S-wave scattering length \cite{Dover} and the
$\pi N \rightarrow K\Lambda$ data of Ref. \cite{Jones} this ratio can be estimated
to be about 9:1, where the main uncertainty is due to the off-shell
extrapolations. This suggests that the $K$ exchange should be the
dominant $\Lambda$ production mechanism. 
For $\Sigma^0$ threshold production, however, the situation is reversed.
Here the $K$ exchange will be strongly suppressed if one assumes 
the SU(6) ratio for the pertinent coupling constants, whereas the
pion-exchange contribution will be somewhat enhanced according to eq. (4).
Therefore the $\Lambda$/$\Sigma^0$ production ratio will be 
given essentially by the ratio of the $K$-exchange in $\Lambda$ production to 
$\pi$-exchange in $\Sigma^0$ production which -- combining the numbers given 
above -- amounts to $R_{pp}(0) \approx 9 \times 0.4 \approx 4$.  
Clearly there is still a discrepancy of a factor of about 7 between this
estimation and the empirical ratio.
 
A large value of $R_{pp}(Q)$ could be obtained if the low energy $\Lambda N$
interaction were attractive and the $\Sigma N$ repulsive. However, in addition
to being at variance with other data, the smallness of the value of
$\epsilon(\Sigma^0p)$ given in eq.~(3) can only be understood for an attractive
interaction.

The COSY-11 acceptance for $pp\to pK^+\Lambda$ near the $\Sigma$ threshold is
poor in the region expected to be strongly affected by the extra two-step
contribution {\it via} the $\Sigma$ hyperon. However, the TOF collaboration at 
COSY has measured the 
exclusive $pp\to pK^+\Lambda$ production at beam momenta of 2.50~GeV/c and
2.75~GeV/c~\cite{TOF}, and the $pK^+\Sigma^0$ threshold lies between these two
values.  The distribution of $\Lambda p$ masses from the higher momentum data 
shows clear evidence for an attractive final state interaction but, in addition,
at an invariant mass $M(\Lambda p) = (2129\pm 2)$~MeV/c$^2$, there is an
isolated point which is high compared to the phase-space distribution.
Since $m(\Sigma) + m(N) \approx  2130$~MeV/c$^2$, we suggest that this is evidence
for $\Sigma N \to \Lambda p$ conversion in this reaction.  Given that the TOF
binning was about 8~MeV/c$^2$, the observation of an excess in a single bin is
completely consistent with the $K^-$-deuterium data~\cite{Tan1}. It seems likely
that the TOF peak is a genuine effect but the $pK^+\Lambda$ production cross 
section should be remeasured just above the $pK\Sigma$ threshold with high 
precision and good statistics. Combining this with the COSY-11 results would 
allow one to deduce $\Sigma N\ \to \Lambda p$ transition parameters.

In conclusion, the experimental data of the near threshold production 
in the associated strange\-ness process favour the 
$\Lambda$ over the $\Sigma^0$ cross section by a factor of about 28. 
A pure kaon exchange could reproduce the presented measurements 
of the \mbox{$\sigma_{T} (pK^+\Lambda)/\sigma_{T}(pK^+\Sigma^0)$} ratio
provided that the SU(6) value is taken for the ratio of the $\Lambda N K^+$ and
$\Sigma N K^+$ coupling constants. However, any reasonable contribution from
pion exchange destroys this agreement and other explanations must be
sought. There is much experimental evidence to suggest that the effect is due
to the produced $\Sigma$'s being converted into $\Lambda$'s through a
$\Sigma N \to \Lambda p$ transition in the final state. This would give at least
one natural
explanation of the observed ratio. Further experimental data which allow one to
decouple both spin and isospin observables are essentially needed. A quantitative 
understanding of the phenomenon is very important since it is precisely in this 
coupling that existing nucleon-hyperon models \cite{Mae89, Holzenkamp} deviate most
strongly.\\
\\
\\
\\
We have had many helpful discussions with and very valuable hints from
A.~Kudryavtsev (ITEP-Moscow); thank you Sasha. We appreciate the work provided 
by the COSY operating team and wish to thank them for the good cooperation and
for delivering the excellent proton beam. The research project was supported by
the BMBF (06MS881I), the Polish Committee for Scientific Research 
(2P03B-047-13), and the Bilateral Cooperation between Germany and Poland 
represented by the Internationales B\"uro DLR for the BMBF (PL-N-108-95). The 
collaboration partners from the Westf\"alische Wilhelms-University of M\"unster 
and the Jagellonian University of Cracow as well as one of the authors (C.H.)
appreciate the support provided by the FFE-grant (4126606, 41266654, 41324880,
respectively) from the Forschungszentrum J\"ulich. One of the authors (C.W.) 
benefitted from a consultancy contract at the Forschungszentrum J\"ulich, 
where this work was carried out.

\end{document}